\documentclass[12pt]{article}
\usepackage{arxiv}
\usepackage{hyperref}
\usepackage{orcidlink}
\usepackage{academicons}
\usepackage{xcolor}
\definecolor{orcidlogocol}{HTML}{A6CE39}  

\usepackage[numbers,sort&compress]{natbib}
\setcitestyle{numbers,square}

\usepackage[T1]{fontenc}
\usepackage[utf8]{inputenc}
\usepackage{newtxtext,newtxmath}  
\usepackage{microtype}            
\usepackage{enumitem}
\usepackage{cuted}
\usepackage{caption}

\usepackage[font=small,labelfont={bf,it},textfont=it]{caption}
\usepackage{capt-of}   

\usepackage{hyperref}
\usepackage{url}
\usepackage{booktabs}
\usepackage{amsfonts}
\usepackage{nicefrac}
\usepackage{graphicx}
\usepackage{doi}

\title{Future-Back Threat Modeling: A Foresight-Driven Security Framework}

\author{
  Vu Van Than, CISSP\thanks{Email: vuvanthancnc@gmail.com}\\
  Head of Cyber Security, Independent Researcher\\
  Ha Noi, Viet Nam\\
  ORCID: \href{https://orcid.org/0009-0001-7561-5304}{0009-0001-7561-5304}
}

\hypersetup{
pdftitle={A template for the arxiv style},
pdfsubject={q-bio.NC, q-bio.QM},
pdfauthor={Vu Van Than},
pdfkeywords={Future-Back Threat Modeling, Cyber Threat Foresight, More},
}

\begin{document}
\maketitle
\begin{abstract}

Traditional threat modeling remains largely grounded in known tactics, techniques, and procedures (TTPs), current system conditions, and evidence from past incidents, while strategic foresight methods often remain disconnected from the operational and architectural artifacts used to test security assumptions. This creates an important gap: significant cyber risk may arise not only from what organizations already know, but also from what they implicitly assume to be safe, overlook, or have not yet subjected to evidence.

To address this gap, this paper develops Future-Back Threat Modeling (FBTM), a foresight-driven and assumption-centered security framework. FBTM uses plausible future business losses as backcasting seeds to expose the safety assumptions that make those loss paths appear controlled today. Threat-intelligence and foresight inputs provide contextual material for examining those assumptions, while outside-in adversary relevance and inside-out system applicability are used to formulate organization-specific, testable scenarios.

Rather than seeking to predict adversary choices with certainty, FBTM uses targeted testing and evidence collection to challenge explicit assumptions and update analytical judgments about relevance, confidence, applicability, and priority. Evidence may support, refine, or contradict an existing assumption, and unresolved questions can generate new or revised intelligence requirements for subsequent cycles. These judgments are then translated through governance and executive decision processes into operational responses and learning.

In this way, FBTM connects foresight, threat modeling, evidence, and governance within a reflexive learning cycle. Its objective is not to eliminate uncertainty, but to make future-oriented security assumptions explicit, testable, and sufficiently decision-ready for organizations to act while meaningful choices remain.

\end{abstract}


As first articulated in ISC2 Insight Future-Back Threat Modeling \cite{isc22025futureback}, the Future-Back approach integrates epistemic foresight into cyber threat modeling, aligning with ISACA’s revisited perspective on systemic assurance \cite{isaca2025threatmodelingwhitepaper}. Collier (2020) further demonstrated how backcasting can reveal systemic weaknesses within cyber threat intelligence ecosystems, supporting its epistemic role in foresight-driven security \cite{collier2020backcasting}.

\section{Introduction -- The Limits of Past-Forward Security}

\subsection{Problem Context}

Recent industry threat-intelligence reports from CrowdStrike, ENISA, and Mandiant show that adversaries are continuously updating their tactics, techniques, and procedures (TTPs) in ways that emphasise stealth and defence evasion across multiple intrusion vectors \cite{crowdstrike2025gtr,enisa2025tl,mandiant2024mtrends}. CrowdStrike's \textit{Global Threat Report 2025} notes that 2024 saw the emergence of threat actors that ``exclusively target cloud environments with unique, cloud-specific skill sets'' and that adversaries ``strengthened their emphasis on defense evasion in cloud environments, \textbf{adopting stealth-oriented tactics and tools for initial access and credential access}'' \cite{crowdstrike2025gtr}.

ENISA's \textit{Threat Landscape 2025} similarly reports that ``\textbf{phishing continued to be the primary method for initial intrusion}'' and highlights newly observed \textbf{ClickFix-style scams} that ``\textbf{appeared during the reporting period},'' as well as the ClearFake campaign, described as ``\textbf{another innovative technique}'' that weaponises compromised WordPress sites to distribute infostealers via fake CAPTCHA and verification prompts \cite{enisa2025tl}. 

Mandiant's \textit{M-Trends 2024} documents how, after Microsoft blocked Office macros by default, attackers adopted ``\textbf{multiple types of phishing payloads and techniques that fell outside of historical norms},'' including ``\textbf{code obfuscation, remote payload hosting, placing dropper scripts within archive files, and bypassing email filtering controls},'' which Mandiant categorises under ``\textbf{Malware Delivery: Old, Borrowed, and New Techniques}'' \cite{mandiant2024mtrends}. 

Taken together, these observations provide converging evidence that adversarial behavior is not static. Attackers adapt to changes in defensive controls, technology architectures, and opportunities for exploitation. This creates a persistent challenge for security practices that rely predominantly on current system states, previously observed behaviors, or established control assumptions.

Threat-modeling methods such as STRIDE and PASTA provide structured ways to identify threats, attack paths, and mitigation requirements, while broader cybersecurity governance and control frameworks such as NIST CSF and ISO 27001 serve different but complementary purposes in organizing cybersecurity outcomes, controls, and risk-management practices. These approaches remain essential; however, they do not by themselves provide an explicit mechanism for working backward from a plausible future business loss to identify and challenge the safety assumptions that make that loss appear controlled today.

The resulting gap is therefore not simply a lack of threat information. Organizations may possess extensive threat intelligence, mature controls, and established risk frameworks while still relying on assumptions that have not been made explicit or tested against plausible future conditions. The challenge is to connect foresight about emerging change with organization-specific assumptions, system context, and evidence in a form that can influence decisions before those assumptions are invalidated by an incident.

Future-Back Threat Modeling (FBTM) is motivated by this gap. Rather than replacing existing threat-modeling methods or cybersecurity governance frameworks, it introduces a complementary Future-Back reasoning process for exposing, contextualizing, and testing the assumptions on which present confidence in security depends.

\begin{table*}[ht]
\centering
\caption{Comparative Mapping of FBTM Emphases Against Major Frameworks and Techniques}
\small
\setlength{\tabcolsep}{4pt}
\renewcommand{\arraystretch}{1.15}

\begin{tabular}{|p{4cm}|p{2.5cm}|p{2.5cm}|p{2.5cm}|p{2.7cm}|p{2.5cm}|}
\hline

\textbf{FBTM Comparison Dimensions} &
\textbf{NIST CSF 2.0} &
\textbf{COBIT} &
\textbf{ISO 27001:2022} &
\textbf{MITRE (ATT\&CK / Engage)} &
\textbf{Honeypots} \\
\hline

Backcasting foresight to expose and challenge future-oriented security assumptions &
\textit{Not explicitly addressed as a primary mechanism} &
\textit{Not explicitly addressed as a primary mechanism} &
\textit{Not explicitly addressed as a primary mechanism} &
\textit{Not explicitly addressed as a primary mechanism} &
\textit{Not explicitly addressed as a primary mechanism} \\
\hline

Understanding the adversary and ourselves through contextualization &
Cybersecurity outcomes, detection, and risk context &
Governance alignment and risk/control objectives &
Risk treatment, controls, and organizational context &
Adversary behavior and TTP mapping &
Observed interaction and adversary telemetry \\
\hline

Targeted testing and strategic sensors as evidence-generating mechanisms for assumption testing &
Partial: detection, monitoring, and control assessment &
Control assurance and testing &
Partial: control monitoring and assurance &
Adversary engagement and behavior-informed testing &
Directly applicable as strategic sensors and deception mechanisms \\
\hline

Anticipating and testing plausible adversary behavior &
Threat-informed detection and risk analysis &
Risk scenario analysis &
Threat intelligence and vulnerability/risk management &
Tactic chaining and adversary behavior modeling &
Observed behavior can inform and challenge threat assumptions \\
\hline

\end{tabular}

\label{tab:framework_comparison}
\end{table*}

\subsection{Theoretical Gap}

Strategic foresight methods, including scenario exploration and backcasting, provide structured ways to reason about plausible long-term change and to question present assumptions through alternative future states. Their strength lies in expanding the range of conditions that decision makers are prepared to consider. However, foresight outputs do not automatically become engineering-ready security artifacts: a plausible future scenario does not by itself identify which current safety assumptions should be examined, where relevant adversary behavior could intersect with a specific environment, or what evidence would justify changing a security judgment.

Conversely, threat modeling, cyber threat intelligence (CTI), security testing, and cybersecurity governance frameworks provide increasingly structured ways to characterize adversaries, systems, exposures, controls, and risk. These capabilities are complementary and already support important parts of security analysis. Their availability, however, does not by itself establish a disciplined path from a plausible future business loss to explicit safety assumptions, organization-specific test scenarios, targeted evidence, and governed decisions.

The theoretical gap addressed by Future-Back Threat Modeling therefore lies not in the absence of foresight, threat intelligence, threat modeling, or security validation as individual practices. Rather, it lies in their orchestration around explicit and challengeable safety assumptions. FBTM connects future-oriented reasoning with adversary relevance, system applicability, targeted evidence, and decision-making through a reflexive learning cycle.

In this formulation, foresight is not used to predict an adversary's next action with certainty. Instead, it establishes plausible future conditions under which current beliefs about security can be examined. Threat intelligence and system context then help determine which adversary behaviors are relevant and locally applicable, transforming those conditions into testable scenarios. Evidence from targeted testing and observation can subsequently support, refine, or contradict the current judgment, while unresolved questions can inform subsequent intelligence requirements and future analytical cycles.

\subsection{Purpose and Contribution}

Future-Back Threat Modeling (FBTM) addresses the theoretical gap identified above by integrating foresight into the operational reasoning of threat modeling without treating foresight as a mechanism for deterministic prediction. Its purpose is to make future-oriented security assumptions explicit, place them within relevant adversary and system context, expose them to evidence, and translate the resulting judgments into governed decisions and organizational learning.

The central unit of analysis in FBTM is the safety assumption: an explicit proposition about the conditions under which a system, control, dependency, process, or organizational capability is expected to remain sufficiently secure. Controls and policies are therefore not themselves hypotheses. Rather, claims about their expected behavior or effectiveness under specified conditions are treated as provisional and testable propositions.

FBTM operationalizes this reasoning through a Future-Back sequence. A plausible future business loss or disruption provides the backcasting seed from which relevant safety assumptions are identified. Threat intelligence and foresight inputs then help contextualize those assumptions, while adversary relevance and system applicability are examined to determine the preconditions and organization-specific scenarios that warrant testing. Targeted testing and observation subsequently produce evidence that may support, refine, or contradict the current assumption.

The contribution of FBTM is therefore not a new threat taxonomy, adversary-behavior model, or control framework. Rather, it is an assumption-centered orchestration mechanism that connects:

Future Loss→Safety Assumption→Contextualized Scenario→Targeted Evidence→Analytical Judgment→Governed Decision→Learning.

This sequence makes explicit how foresight can be translated into testable security questions and how evidence can change confidence before uncertainty collapses available choices. FBTM therefore seeks not to prove that systems are safe, nor to predict adversary behavior with certainty, but to improve the quality, timing, and traceability of security judgments under adversarial uncertainty.


\section{Related Work -- Where FBTM Extends Existing Thinking}
\label{sec:relatedwork}

Traditional threat-modeling methods such as STRIDE and PASTA \cite{pasta_rcTM}, 
together with broader cybersecurity frameworks such as NIST CSF, 
guide teams to identify known threats and map controls to reduce risk 
\cite{microsoft2005stride,nistcsf2024}. 

At the field level, Xiong and Lagerstr\"{o}m characterize threat modeling
as a domain that lacks common ground and, based on their systematic
literature review, report limited assurance regarding the validation
of the reviewed work~\cite{xiong2019threatmodeling}.

However, these approaches are primarily present- or past-informed: 
they generally begin with current systems, known attacker behavior, 
or established risk conditions rather than working backward from plausible future states.

In contrast, many foresight-oriented approaches focus on constructing plausible future scenarios 
but seldom examine whether the assumptions underlying those scenarios would hold under adversarial 
or unexpected conditions. They help organizations imagine what might happen, 
but they do not necessarily translate those futures into testable security hypotheses 
or mechanisms for validating safety assumptions.

Similarly, backcasting in cyber intelligence \citep{collier2020backcasting} 
helps counter hype and clarify future perspectives, 
but does not by itself connect those insights to technical validation.

Future-Back Threat Modeling (FBTM) bridges these worlds: 
it begins with envisioned futures (\textit{foresight}), 
challenges the assumptions that make us feel safe today (\textit{epistemic testing}), 
and converts insights into specific technical and governance artefacts 
(\textit{flags} $\rightarrow$ \textit{gates} $\rightarrow$ \textit{controls}).

In short, FBTM translates imagination into evidence.

\raggedbottom

\section{Foundational Principles of Future-Back Threat Modeling}
FBTM rests on four interlocking principles that redefine how organizations think about assurance, foresight, and learning.
It moves beyond control-centric threat modeling toward a reflexive discipline that learns by challenging its own safety
assumptions through evidence.

\subsection{Core Principles of Future-Back Threat Modeling (FBTM)}
The Future-Back Threat Modeling (FBTM) framework is grounded in four interrelated epistemic principles that
differentiate it from traditional threat modeling. Each principle reframes how security foresight, validation, and
governance interact across temporal and cognitive boundaries.

\begin{enumerate}
\setlength{\itemsep}{0.7em}
\setlength{\parskip}{0pt}
\setlength{\parsep}{0pt}
\setlength{\topsep}{0.5em}

\item \textbf{Temporal Inversion.}
FBTM starts from the future and works backward rather than projecting threats forward
from the present. It treats a future breach or disruption as if it has already occurred, then reverses the timeline
to uncover the fragile assumptions that made it possible. Assumptions---not predefined threats---form the
central unit of analysis.

\item \textbf{Epistemic Stress-Testing.}
Safety assumptions, including claims about the expected effectiveness of controls or policies, are treated as
testable hypotheses that can be challenged by evidence. Security confidence is calibrated through evidence that
may support, refine, or contradict an assumption, with deliberate attention to potentially disconfirming evidence.
Deception environments, controlled experiments, anomaly detection, and other targeted testing mechanisms can
help expose conditions under which current beliefs about safety may no longer hold.

\item \textbf{Flags $\rightarrow$ Gates $\rightarrow$ Controls.}
Future-oriented signals and early indicators (\textit{flags}) identify safety assumptions that warrant examination.
These signals do not move directly to controls; they are first contextualized, tested, and evaluated through evidence.
The resulting analytical judgments then pass through governance and executive verification points (\textit{gates}), which may lead to measurable safeguards (\textit{controls}) or other governed risk decisions.
This chain operationalizes foresight by connecting emerging signals to evidence-informed decisions, controls, and subsequent organizational learning.

\item \textbf{Reflexive Learning and Perspective over Panic.}
Each tested or revised assumption enriches the organization's foresight knowledge base, refining subsequent
assumptions and strengthening adaptive resilience. FBTM promotes perspective over panic---ranking futures by
systemic impact rather than novelty---to ensure foresight drives proportionate and strategic responses instead
of reactive measures.

\end{enumerate}

\begin{center}
\begin{minipage}{\linewidth}
  \centering
  \includegraphics[width=\linewidth,height=.78\textheight,keepaspectratio]{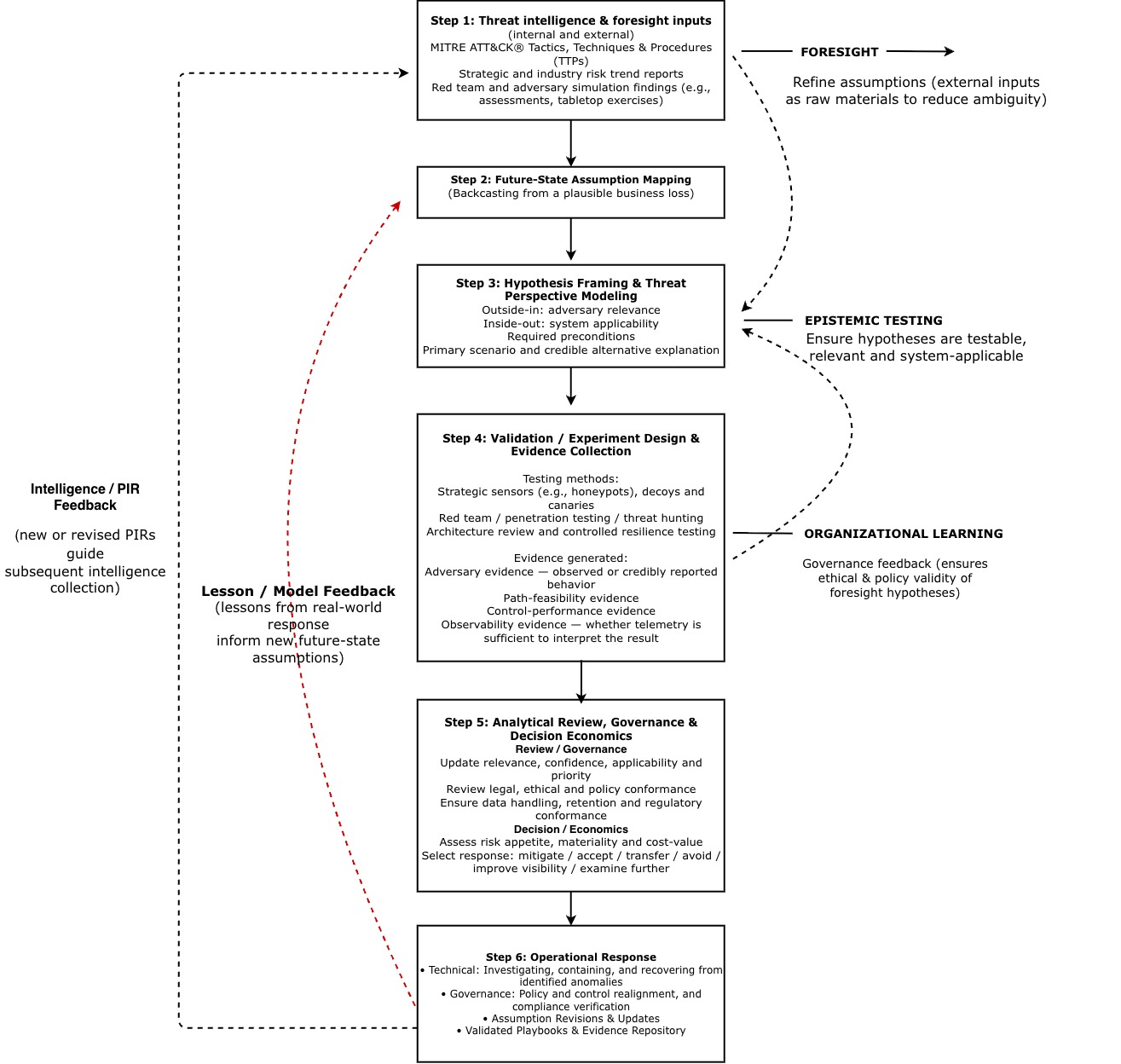}
  \captionof{figure}{Refined workflow of Future-Back Threat Modeling (FBTM).
The framework presents a six-step operationalization of the Future-Back cycle while
clarifying the relationships among foresight, assumption mapping, adversary and system
contextualization, evidence generation, analytical judgment, governed decision-making,
and organizational learning. Feedback loops revise future-state assumptions and generate
new or refined priority intelligence requirements (PIRs) for subsequent cycles.}
  \label{fig:fbtm.jpg}
\end{minipage}
\end{center}

\section{Method Overview -- The Future-Back Reflexive Workflow}
\subsection{Phase I -- Foresight and Assumption Mapping}

This phase inverts the direction of traditional threat modeling by working backward from plausible future states rather than beginning solely with current systems.
The objective is to articulate the beliefs, expectations, and biases that shape an organization's sense of safety, then make them explicit and testable.
This reflects \textbf{Karl Popper}'s emphasis on falsifiability and the importance of exposing assumptions to potentially disconfirming evidence \cite{popper1959falsification},
and aligns with \textbf{John Robinson}'s insight that backcasting reveals how future scenarios depend on present-state assumptions \cite{robinson1990backcasting}.

\subsubsection{Step 1 -- Framing Intelligence Inputs}

Gather diverse foresight and threat-intelligence sources, including adversarial simulations, horizon scans, socio-technical trend reports, and geopolitical signals.
These are not treated as deterministic threat predictions, but as contextual inputs that help frame plausible future conditions and inform subsequent assumption mapping.

\textbf{Output:} a set of relevant foresight and threat-intelligence inputs for Future-Back analysis.

\subsubsection{Step 2 -- Future-State Assumption Mapping}

Using a plausible future business loss or disruption as the backcasting seed, identify the safety assumptions that must hold for the system or organization to remain sufficiently secure.
Classify each assumption according to its relative fragility or robustness.
Each explicit assumption becomes a hypothesis seed---the epistemic unit that drives the next phase.

The following illustrative map combines global foresight with domain-specific future conditions in a hypothetical FinTech scenario to demonstrate how safety assumptions can be identified and prioritized.

\begin{table*}[ht]
\centering
\caption{Illustrative Phase I -- Integrated Assumption Mapping Combining Global and Domain-Specific Foresight.
A plausible future business loss or disruption provides the backcasting seed, while ENISA-derived foresight is used as contextual input for identifying and prioritizing safety assumptions.}
\label{tab:phase1_assumption_mapping}

\small
\setlength{\tabcolsep}{3pt}
\renewcommand{\arraystretch}{1.15}

\begin{tabular}{|p{0.6cm}|p{2.6cm}|p{2.8cm}|p{1.6cm}|p{1.2cm}|p{3cm}|p{5cm}|}
\hline

\textbf{ID} &
\textbf{Foresight Context (ENISA)} &
\textbf{Security Assumption (to test)} &
\textbf{Category} &
\textbf{Fragility} &
\textbf{Early Flags / Indicators} &
\textbf{Business Context \& Stakeholder Impact} \\
\hline

A1 &
Supply-Chain Compromise of Software Dependencies &
Third-party dependencies and open-source components remain trustworthy; no hidden backdoors infiltrate the build process. &
Technical / Organizational &
Fragile &
Unverified commits from unknown contributors; spike in new or unscanned dependencies; advisory alerts from NVD / GitHub. &
\textbf{Global foresight:} ENISA warns of systemic OSS compromise.
\textbf{Domain foresight:} Hypothetical FinTech platform scales its vendor ecosystem substantially by 2040; vendor-risk and DevSecOps teams must maintain continuous dependency and SBOM assurance. \\
\hline

A2 &
Skill Shortages \& Automation Gaps &
Operational assurance will remain stable as AI-enabled automation replaces or augments human analysts. &
Organizational / Sociotechnical &
Fragile &
SOC backlog exceeds baseline; increasing false-positive burden; prolonged hiring or specialist-capability gaps. &
\textbf{Global foresight:} ENISA identifies cyber-talent scarcity as a systemic concern.
\textbf{Domain foresight:} SOC staffing remains constrained while reliance on AI-enabled security tooling increases; HR and the CISO share responsibility for capability resilience. \\
\hline

A3 &
Artificial Intelligence Abuse / Data Poisoning &
AI models and data pipelines remain sufficiently protected against poisoning, manipulation, and integrity degradation. &
Technical / Ethical &
Fragile &
Unexplained model or score drift; increase in adversarial inputs; anomalous model outputs or data-integrity indicators. &
\textbf{Global foresight:} ENISA highlights model manipulation and AI misuse.
\textbf{Domain foresight:} AI-supported fraud detection becomes increasingly material to transaction approval and risk decisions; the CDO and AI governance function become key stakeholders. \\
\hline

A4 &
Advanced Disinformation \& Reputation Attacks &
Information-integrity controls can distinguish sufficiently between authentic and synthetic narratives capable of materially affecting the organization. &
Cognitive / Legal &
Fragile &
Increase in deepfake activity; coordinated bot or influence patterns; abnormal information-manipulation indicators. &
\textbf{Global foresight:} ENISA identifies increasingly sophisticated information manipulation and synthetic-content risks.
\textbf{Domain foresight:} A multinational FinTech organization faces cross-jurisdictional reputational and legal exposure; Corporate Communications and Legal Risk require rapid assessment and response mechanisms. \\
\hline

A5 &
Climate-Driven Infrastructure Disruption &
Critical data-centre and service dependencies remain sufficiently resilient under regional environmental stress and related supply-chain disruption. &
Operational / Strategic &
Robust $\rightarrow$ Fragile &
Regional power instability above baseline; material supply-chain delays; environment-related service degradation or downtime. &
\textbf{Global foresight:} ENISA anticipates increasing climate-related infrastructure and dependency risks.
\textbf{Domain foresight:} Data-centre dependencies in Southeast Asia face increasing thermal, energy, and continuity pressures; CIO and BC/DR leadership must maintain adaptive resilience. \\
\hline

\end{tabular}
\end{table*}

\noindent
This illustrative map combines \textbf{global foresight}---including EU-level systemic threats identified by ENISA \cite{enisa2030foresight}---with
\textbf{domain-specific foresight} derived from the projected future conditions of a hypothetical FinTech organization.
Each assumption represents an explicit and \textit{testable belief} about the organization's future safety conditions.
Fragility ratings indicate where \textbf{global uncertainty} may interact with \textbf{enterprise-level stressors},
helping prioritize assumptions for subsequent examination.

The top three critical fragile assumptions---A1 (Supply Chain Integrity),
A2 (Skill Resilience), and A3 (AI Trustworthiness)---constitute the
\textit{epistemic focus} for Phase~II, where testable hypotheses and targeted tests are constructed.
The resulting evidence subsequently informs the analytical and governance gates in Phase~III.

\section{Phase II -- Epistemic Testing and Validation}
\label{sec:others}

\subsection{Purpose and Rationale}

Each safety assumption identified in Phase~I is restated as an explicit, testable hypothesis, serving as the basis for empirical examination.
Rather than seeking only to confirm that an organization's controls are effective, FBTM deliberately examines conditions under which current safety assumptions may fail.
This reversal---testing ``what might challenge our belief in safety'' rather than only ``what sustains it''---creates an epistemic stress test in which evidence may support, refine, or contradict the current assumption.
The objective is to expose hidden vulnerabilities, weak control expectations, and unexamined dependencies before they contribute to material loss.

\subsection{Step 3 -- Hypothesis Framing and Threat Perspective Modeling}

Following the epistemic mapping of assumptions in Phase~I, this phase introduces a structured, multi-perspective method
to reframe each assumption into a testable hypothesis.
Drawing on established threat-modeling and risk-assessment practices, together with the
\textbf{MITRE ATT\&CK framework}, threat perspectives are examined through three complementary lenses:

\begin{itemize}
    \item \textbf{Attacker-centric (who):} Focusing on adversarial capability, intent, targeting, tactics, techniques, and relevant actor behavior.
    \item \textbf{Asset-centric (what/why):} Identifying high-value targets, critical data flows, identities, privileges, and potential business consequences.
    \item \textbf{Architecture-centric (how):} Understanding vulnerabilities, interdependencies, trust boundaries, controls, and potential attack paths within systemic design.
\end{itemize}

Together, these perspectives connect \textit{outside-in adversary relevance} with
\textit{inside-out system applicability}.

\begin{figure*}[ht]
    \centering
    \includegraphics[width=0.95\textwidth]{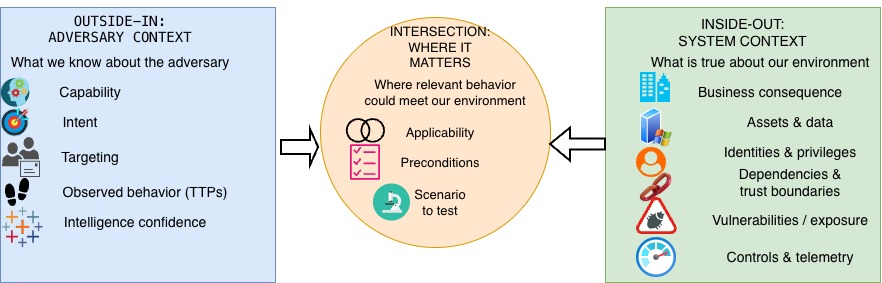}
    \caption{Contextualizing adversary behavior in Future-Back Threat Modeling.
Outside-in adversary context establishes relevance, while inside-out system
context establishes local applicability. Their intersection identifies the
preconditions and organization-specific scenarios that warrant testing.
Author's synthesis informed by NIST SP 800-30 Rev. 1
\cite{nist2012sp80030r1}.}
    \label{fig:contextualization}
\end{figure*}

This three-dimensional approach goes beyond typical ``threat enumeration'' to model
\textit{epistemic exposure}---the limits of what the organization believes to be secure.
It highlights where safety assumptions have not been tested, particularly along high-value operational paths,
dependencies, or architectural controls.

The analysis identifies the required preconditions, formulates the primary organization-specific scenario to be tested,
and, where appropriate, considers at least one credible alternative explanation.
This lightweight use of competing explanations follows established intelligence-analysis practice for reducing premature commitment to a single interpretation when evidence is incomplete or ambiguous\cite{heuer1999psychology}.
This helps prevent externally observed adversary behavior from being treated as automatically applicable to the local environment.

By aligning these perspectives with future-state assumptions, the framework turns each assumption into a testable statement:

\begin{quote}
    \textit{If assumption X is valid, system Y should remain sufficiently secure under future condition Z.}
\end{quote}

Each hypothesis is recorded in a \textbf{hypothesis register} including:

\begin{enumerate}
    \item The original safety assumption;
    \item Its related future-state condition;
    \item The relevant adversary and system context;
    \item The required preconditions and primary scenario;
    \item The criteria or evidence that would support, refine, or contradict the assumption; and
    \item Early signs that may indicate epistemic failure or insufficient observability.
\end{enumerate}

This approach is conceptually inspired by \textbf{ISACA (2025), \textit{Threat Modeling Revisited}},
which suggests a five-stage cycle: goal setting, ecosystem mapping, threat prioritization, mitigation planning, and iterative validation.
However, FBTM inverts the analytical starting point: instead of beginning solely with enumerated threats and moving toward controls,
it begins with future-oriented safety assumptions and works backward to determine which adversary behaviors,
system conditions, and control expectations warrant examination.
Adversarial simulation and targeted evidence collection are then used to challenge those assumptions before mitigation decisions are made.
This \textit{epistemic inversion} distinguishes FBTM from conventional present-oriented threat analysis.

\subsection{Step 4 -- Validation, Experiment Design, and Evidence Collection}

The goal is not to prove that existing controls are safe, but to identify and challenge the safety assumptions that underpin confidence in them.
FBTM therefore selects testing and evidence-collection mechanisms according to the hypothesis under examination.

Relevant mechanisms may include:

\begin{itemize}
    \item \textbf{Strategic sensors and deception environments}---including honeypots, decoy credentials, canary assets, or synthetic data \cite{isc22026honeypot};
    \item \textbf{Red-team exercises and adversary emulation};
    \item \textbf{Penetration testing and targeted threat hunting};
    \item \textbf{Architecture, configuration, and control reviews}; and
    \item \textbf{Controlled resilience or chaos experiments}, where appropriate.
\end{itemize}

Each test is assessed across three cross-cutting evaluation dimensions:
\textbf{technical robustness}, \textbf{organizational responsiveness}, and
\textbf{ethical proportionality}.
Within these dimensions, evidence is distinguished across four complementary categories:

\begin{itemize}
    \item \textbf{Adversary evidence:} Whether relevant behavior has been directly observed or credibly reported;
    \item \textbf{Path-feasibility evidence:} Whether the technical and organizational preconditions required for the modeled scenario exist;
    \item \textbf{Control-performance evidence:} How relevant preventive, detective, or responsive controls behave under the examined conditions; and
    \item \textbf{Observability evidence:} Whether available telemetry is sufficient to interpret the result reliably.
\end{itemize}

\begin{figure}[htbp]
    \centering
    \includegraphics[width=0.95\textwidth]{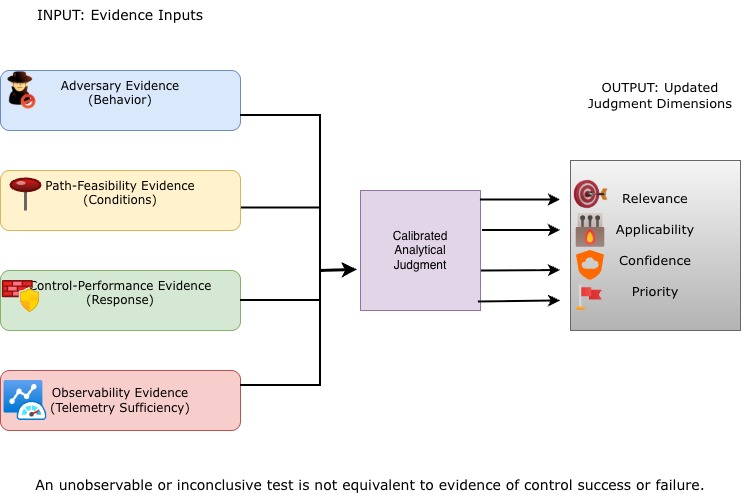}
    \caption{Four-category evidence flow for calibrated analytical judgment in
    Future-Back Threat Modeling (FBTM). Evidence from adversary behavior,
    path feasibility, control performance, and observability informs analytical
    judgment and updates relevance, applicability, confidence, and priority.}
    \label{fig:fbtm-evidence-flow}
\end{figure}

A failed test and an unobservable test are therefore not treated as equivalent
findings. Instead of producing binary proof or probabilistic certainty, the
analysis yields calibrated confidence judgments indicating how strongly the
available evidence supports, refines, or contradicts the current assumption.

A failed test and an unobservable test are therefore not treated as equivalent findings.
Instead of producing binary proof or probabilistic certainty, the analysis yields calibrated confidence judgments indicating how strongly the available evidence supports, refines, or contradicts the current assumption.

This preserves the epistemic focus of FBTM: validation is not about proving systems safe, but about making safety assumptions explicit and exposing them to evidence capable of changing the current judgment.
Accordingly, organizational learning is reflected not simply in how many assumptions fail, but in how systematically evidence is used to revise assumptions, controls, and subsequent decisions.

\section{Phase III -- Governance, Executive Perception, and Post-Modeling Action}

At this stage, the focus shifts from technical and analytical testing to how the resulting evidence reshapes governance and decision-making.
Future-Back Threat Modeling (FBTM) does not replace frameworks such as NIST CSF or RMF; rather, it can operate alongside them, turning tested assumptions and evidence into learning and decision inputs.
When a safety assumption is challenged, the point is not blame or proof---it is translation: how does this new knowledge alter the way we govern, prioritize, and act?

\subsection{Step 5a -- Analytical Review, Governance and Legal Gate}

After each epistemic test, the available evidence is first reviewed to update analytical judgments about
\textbf{relevance}, \textbf{confidence}, \textbf{applicability}, and \textbf{priority}.
These judgments distinguish what is known about the adversary from what is applicable to the organization's environment and how strongly the available evidence supports the current assessment.

The results are then reviewed through a governance gate that can be aligned with the Assess--Authorize--Monitor logic of the NIST Risk Management Framework.
This is not simply a bureaucratic stop; it is where analytical evidence becomes accountable risk information.

Three governance lenses guide this review:

\begin{enumerate}
    \item \textbf{Legal and ethical alignment}---ensuring that tests such as honeypots, deception environments, or adversary simulations comply with applicable data-handling, privacy, and ethical obligations.

    \item \textbf{Policy and control connection}---mapping what has been learned back to existing policies and control frameworks, including NIST SP 800-53 \cite{nist2020sp80053}
and ISO/IEC 27001 \cite{iso27001_2022}.

    \item \textbf{Risk appetite review}---asking whether the revised judgment changes what the organization considers acceptable risk and whether additional action or evidence is required.
\end{enumerate}

When done well, this step turns epistemic evidence into governed risk artifacts such as updated control objectives, risk decisions, monitoring requirements, or revised analytical priorities.

\subsection{Step 5b -- Executive Risk Perception and Decision Economics}

In practice, the hardest part of foresight is not data---it is translation.
Executives make decisions in terms of trade-offs, timing, materiality, business constraints, and uncertainty.
The goal of this step is to bridge updated analytical judgments and evidence with executive decision-making.

When a future-oriented risk becomes sufficiently relevant and applicable, decision makers review the evidence against organizational risk appetite, business constraints, materiality, and the expected cost and value of available responses.

Possible responses include:

\begin{itemize}
    \item \textbf{Mitigate:} Adjust defenses, controls, detections, playbooks, or investment priorities.

    \item \textbf{Accept:} Explicitly record why the remaining uncertainty or exposure is tolerable and who owns that decision.

    \item \textbf{Transfer or Avoid:} Shift part of the exposure through partners, insurers, contractual arrangements, or changes to the business activity.

    \item \textbf{Improve Visibility:} Strengthen telemetry, monitoring, or collection where the available evidence is insufficient to support a reliable judgment.

    \item \textbf{Examine Further:} Conduct additional analysis, testing, or intelligence collection where material uncertainty remains.
\end{itemize}

The objective is not to convert uncertainty into artificial certainty, but to make the uncertainty sufficiently explicit and decision-ready for accountable action.

\subsection{Step 6 -- Operational Response and Learning}

Once governance and executive decisions are made, the cycle loops back into operations.
Here, FBTM connects with incident-response and cybersecurity risk-management practices, including NIST SP 800-61 Rev. 3 \cite{nist2025sp80061r3}, while treating operational events not only as situations to contain or recover from, but also as sources of evidence about the assumptions on which current security judgments depend.

When controls fail, incidents occur, or systems behave unexpectedly:

\begin{itemize}
    \item Relevant safety assumptions are made explicit and reviewed.

    \item Detection and analysis examine whether the available evidence supports, refines, or contradicts those assumptions.

    \item Containment, recovery, and control changes reveal how design and operational assumptions behave under real-world conditions.

    \item Lessons from the response are used to revise future-state assumptions, controls, playbooks, and subsequent analytical priorities.
\end{itemize}

Validated findings and lessons are retained in the organization's evidence and learning repositories, creating a cumulative record of how assumptions and judgments change over time.
This turns operational response from a terminal activity into part of a continuing assurance and learning process.

Unresolved questions may also generate new or refined \textbf{priority intelligence requirements (PIRs)}, providing a complementary feedback path into subsequent threat-intelligence collection and Future-Back cycles.

\subsection{Closing the Loop}

The mark of a mature organization is not simply how few incidents occur, but how systematically it can learn from evidence and revise assumptions before uncertainty removes meaningful choices.

In FBTM, the loop---

\[
\textit{Assumption}
\rightarrow
\textit{Hypothesis}
\rightarrow
\textit{Evidence}
\rightarrow
\textit{Judgment}
\rightarrow
\textit{Governed Decision}
\rightarrow
\textit{Response}
\rightarrow
\textit{Learning}
\]

---keeps foresight connected to day-to-day security practice.
It is not about predicting the future with certainty; it is about making uncertainty explicit, testable, and sufficiently decision-ready to support timely action.

\section{Phase IV -- Synthesis and Strategic Implications}

The closing phase of Future-Back Threat Modeling (FBTM) moves from analysis to synthesis---from epistemic testing to strategic sense-making.
While the earlier phases examined how safety assumptions withstand or change under evidence, this stage asks how those lessons reshape foresight, policy, and organizational learning.

FBTM's contribution is not to introduce another risk framework, but to re-architect the flow of knowledge between foresight and control.
Instead of treating security as a compliance perimeter, it reframes it as an adaptive system of inquiry.
Each epistemic cycle---from assumption to evidence and revised judgment---adds reflexive intelligence to the organization's risk posture.

Three strategic implications emerge:

\begin{enumerate}

    \item \textbf{From Compliance to Cognition:}
    FBTM extends assurance beyond checklist-based verification toward continuous examination of the assumptions that support confidence in security.
    Claims about the expected behavior and effectiveness of controls become testable propositions rather than unquestioned beliefs.
    This encourages continuous reasoning about what the organization believes to be ``safe'' and what evidence would justify revising that judgment.

    \item \textbf{Foresight as a Decision Function:}
    By integrating future-state mapping with analytical and governance gates, FBTM embeds foresight directly into executive decision cycles.
    The focus shifts from predicting discrete threats to examining plausible future conditions, testing relevant assumptions, and preparing proportionate responses to uncertainty.

    \item \textbf{Learning as Assurance:}
    In the long term, epistemic learning---the ability to revise assumptions and reduce unjustified confidence in response to evidence---becomes an important dimension of resilience.
    The most adaptive organizations are not those that assume perfect protection, but those that systematically learn from challenged assumptions and incorporate those lessons into subsequent decisions.

\end{enumerate}

In this sense, FBTM does not stand outside existing standards and frameworks such as NIST CSF \cite{nistcsf2024}
or ISO/IEC 27001 \cite{iso27001_2022}.
Rather, it can complement them by providing an assumption-centered learning process that connects foresight, evidence, governance, and operational decision-making.

\section{Discussion and Limitations}

Like all foresight-driven frameworks, FBTM depends on the \textbf{quality of human judgment} and the contextual richness of its inputs.
While it systematizes how organizations expose and examine safety assumptions, it cannot guarantee that those assumptions are selected, contextualized, or tested without bias.
Analytical judgments about adversary relevance, system applicability, and evidence therefore remain subject to incomplete information, competing interpretations, cognitive bias, and organizational blind spots \cite{heuer1999psychology}.
Moreover, the framework's epistemic focus---making assumptions explicit and exposing them to potentially disconfirming evidence---may challenge conventional assurance cultures that prefer certainty and documentation over doubt and discovery.

A second limitation concerns \textbf{evidence quality and observability}.

Prior work on cyberattack modelling has noted that attack models are
incomplete and may require continual updating as the threat landscape
changes, while also proposing annotation factors such as confidence
in information and variance in observations~\cite{happa2017properties}.

An apparent absence of adversary activity or control failure cannot always be interpreted as evidence of safety when telemetry is incomplete or the relevant conditions were not observable.
Accordingly, FBTM distinguishes between a failed test and an unobservable or inconclusive test, treating uncertainty about evidence as part of the analytical judgment rather than as proof of control effectiveness.

A third limitation lies in \textbf{scalability}.
Applying epistemic stress-testing across a large enterprise demands tooling, telemetry, assumption management, and cross-domain coordination.
Automating parts of this process---for example, through AI-augmented horizon scanning, dynamic assumption registries, or evidence management---remains an open area for exploration, but automation does not remove the need for accountable human judgment.

Finally, \textbf{empirical validation} will require longitudinal and comparative studies.
Pilot implementations in critical infrastructure, financial services, or other complex environments could examine whether assumption-centered learning produces measurable improvements in decision quality, control adaptation, or organizational resilience relative to existing practices.

Despite these constraints, FBTM provides a \textit{conceptual and operational bridge} between strategic foresight and security practice.
Its epistemic emphasis is not on falsification for its own sake, but on making assumptions explicit, exposing them to evidence, and revising security judgments when the available evidence warrants change.

\section{Conclusion}

Future-Back Threat Modeling (FBTM) reframes cybersecurity assurance as a discipline of learning under uncertainty.
By moving from foresight to explicit assumptions, targeted testing, and evidence-calibrated judgment, and from compliance to cognition, it enables organizations to govern risk as a living process rather than a static state.

In doing so, FBTM reframes a central question of threat modeling:
instead of asking only ``what could attack us,'' it also asks ``what are we assuming to be safe, and what evidence would make us reconsider that judgment?''
Each meaningfully tested assumption can produce an epistemic gain---whether the evidence supports, refines, or contradicts the current judgment---providing a new data point in how the organization perceives and adapts to future conditions.

Ultimately, the strength of an organization's security lies not in eliminating uncertainty or predicting every threat, but in how effectively it learns while meaningful choices remain.
FBTM provides a language, a method, and a mindset for making that learning deliberate, evidence-informed, and decision-ready.

\vspace{1em}
\noindent\textit{Acknowledgments.}
The author thanks the ISC2 Insights editorial team and foresight practitioners for their early dialogue on epistemic threat modeling.

\section{Case Study}

\textbf{\textit{Applying the Future-Back Threat Modeling Framework to CVE-2025-64446 Using a Simulated FortiWeb Control-Plane Honeypot.}}

This case study provides a worked demonstration of how FBTM translates an explicit safety assumption into an organization-specific scenario, a targeted evidence-collection mechanism, and an updated analytical judgment.
It is intended to illustrate the assumption--hypothesis--evidence sequence of the framework rather than to demonstrate predictive capability or provide empirical validation of FBTM as a whole.

\subsection{Purpose and Future-Back Assumption}

The case begins from a plausible future condition in which an Internet-exposed security-control management interface becomes a path to material compromise.
The relevant safety assumption is:

\begin{quote}
\textit{The exposed WAF control-plane interface remains sufficiently protected against exploitation attempts targeting its management API.}
\end{quote}

This assumption provides the hypothesis seed for examining whether adversary behavior associated with exploitation of FortiWeb control-plane surfaces is relevant to the modeled exposure and whether targeted telemetry can provide evidence capable of changing the current judgment.

The demonstration does not test the broader control-effectiveness claim in full; it tests a subordinate proposition necessary to evaluate it—whether exploit-oriented behavior relevant to the modeled vulnerability can reach the exposed management surface under the defined conditions.

\subsection{Experimental Environment}

\begin{center}
\begin{minipage}{0.95\linewidth}
    \centering
    \captionof{table}{System Components Used in the Future-Back Threat Modeling Case Study for CVE-2025-64446}
    \label{table:futureback-cve64446-components}
    \small
    \begin{tabular}{|l|l|l|l|l|}
    \hline
    \textbf{Component} & \textbf{Hardware} & \textbf{OS} & \textbf{Software / Services} & \textbf{Purpose} \\
    \hline
    Honeypot   & 2 vCPU/2 GB & Ubuntu 22.04 & DShield, mock API, Python, stunnel4 & WAF exposure \\
    \hline
    API Layer  & Same host   & Ubuntu 22.04 & Flask API, decoys, logging          & API emulation \\
    \hline
    Telemetry  & Same host   & Ubuntu 22.04 & DShield logs, journal, logger       & Traffic capture \\
    \hline
    Network    & Public IPv4 & N/A          & Port 8443, stunnel4                 & Attack surface \\
    \hline
    \end{tabular}
\end{minipage}
\end{center}

\begin{center}
\begin{minipage}{0.95\linewidth}
    \centering
    \includegraphics[width=\linewidth, keepaspectratio]{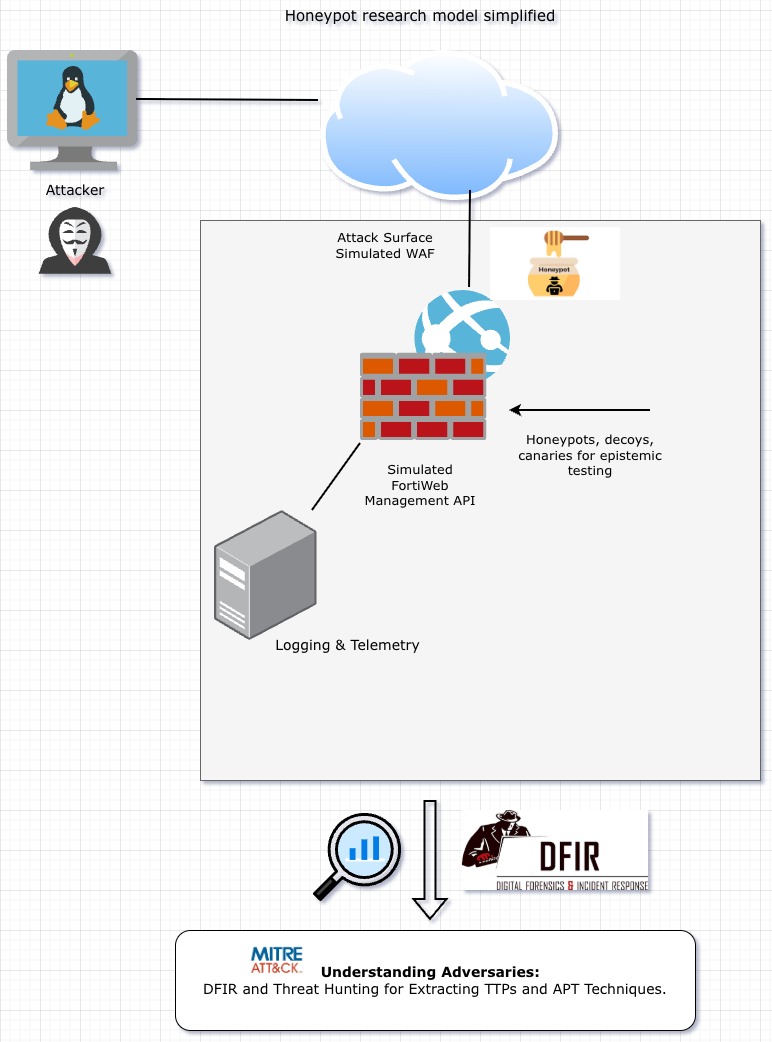}
    \captionof{figure}{Simplified honeypot research model used for the CVE-2025-64446 case study.}
    \label{fig:honeypot-model}
\end{minipage}
\end{center}

A lightweight DShield Honeypot was deployed on an Ubuntu server (2 vCPU, 2~GB RAM) and extended to host a simulated FortiWeb management API.

The experimental environment consists of the following components:

\begin{itemize}
    \item \textbf{Attack Surface}: A simulated WAF reachable on port~8443.
    \item \textbf{Fake FortiWeb API Endpoints}: Non-functional management-plane paths that accept inbound HTTP(S) requests.
    \item \textbf{Honeypot Elements}: Decoys and canary fields used to analyze interaction patterns.
    \item \textbf{Logging and Telemetry}: Full packet inspection and application-level request capture.
\end{itemize}

\subsection{Hypothesis and Threat Context}

The modeled scenario combines \textit{outside-in adversary relevance} with \textit{inside-out system applicability}.

From the outside-in perspective, public reporting indicated exploitation activity associated with CVE-2025-64446 against FortiWeb management-plane surfaces \cite{rapid7_fortiweb_2025}.
From the inside-out perspective, the simulated environment exposed FortiWeb-like management API paths on a public interface, creating the local preconditions required to observe whether similar behavior would reach the modeled surface.

The primary explanation under examination was:

\begin{quote}
\textbf{H1:} The exposed FortiWeb-like management surface will receive exploit-oriented probing consistent with publicly reported CVE-2025-64446 behavior.
\end{quote}

A credible competing explanation was also retained:

\begin{quote}
\textbf{H2:} The observed traffic may reflect generic automated Internet scanning rather than deliberate targeting of the deployed environment.
\end{quote}

The distinction is important because observed behavior can establish relevance without, by itself, establishing adversary identity, intent, or organization-specific targeting.

\subsection{Real-World Telemetry}

After exposing the simulated FortiWeb management API, the honeypot received unsolicited external traffic consistent with exploit-oriented probing associated with CVE-2025-64446, which had been publicly reported as exploited in the wild \cite{rapid7_fortiweb_2025}.

\subsection{Correspondence With Public Reports}

The telemetry captured by the honeypot closely matched indicators independently reported by SANS ISC honeypots \cite{sans_fortiweb_2025}.
Observed requests included:

\begin{itemize}
    \item Access attempts to:
    \begin{quote}
        \texttt{GET /api/v2.0/cmdb/system/admin/../../../../../cgi-bin/fwbcgi}
    \end{quote}

    \item Path-traversal sequences consistent with those documented in public reports;

    \item User-Agent strings commonly associated with automated scanning activity, including Firefox~134.x, Safari/537.36, and Python-requests; and

    \item Repeated probing from external IP ranges, including \texttt{113.x.x.x}.
\end{itemize}

\noindent
\begin{quote}
\small
\textbf{Highlight.}
Example of unsolicited exploit-oriented probing captured by the deployed honeypot.
The requests targeted the simulated FortiWeb management API and matched path-traversal and
\texttt{fwbcgi} patterns reported by SANS ISC.
The experimental architecture is illustrated in Figure~\ref{fig:honeypot-model}.
\end{quote}

\begin{figure}[ht]
    \centering
    \includegraphics[width=\linewidth]{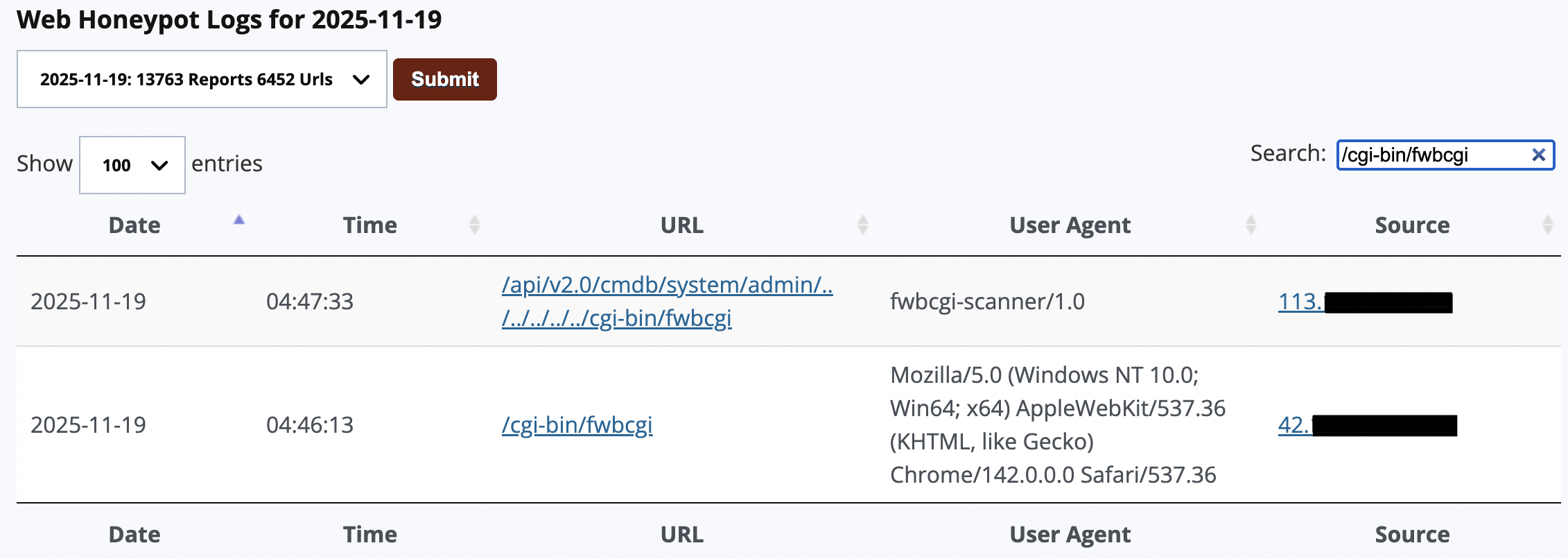}
    \caption{Web honeypot logs for 2025-11-19 showing probing of \texttt{/cgi-bin/fwbcgi}.}
    \label{fig:honeypot-logs}
\end{figure}

\subsection{Honeypot-Specific Observations}

Within the simulated environment, the honeypot recorded:

\begin{itemize}
    \item Full request lines containing the expected \texttt{fwbcgi} exploitation patterns;
    \item Multiple traversal injections (e.g., \texttt{../../../../../});
    \item Enumeration attempts across the simulated API endpoints; and
    \item Probes generated from both Linux- and Windows-based HTTP clients.
\end{itemize}

The correspondence between these observations and independent public reporting provides corroborating evidence for the adversary-behavior component of the modeled scenario.
It does not, however, establish that the deployed environment was deliberately targeted by a specific actor.

\subsection{Evidence Assessment}

Consistent with the FBTM evidence model described in Phase~II, the observations can be separated into four complementary categories:

\begin{itemize}

    \item \textbf{Adversary evidence:}
    Unsolicited requests contained path-traversal and \texttt{fwbcgi} patterns consistent with behavior independently reported in relation to CVE-2025-64446.
    This supports the relevance of the modeled adversary behavior.

    \item \textbf{Path-feasibility evidence:}
    The simulated management surface was publicly reachable, and external requests successfully reached the modeled API paths.
    This demonstrates that the exposure conditions required for probing existed within the experimental environment.
    Because the API was intentionally non-functional, the experiment does not establish exploitability of a genuine FortiWeb appliance.

    \item \textbf{Control-performance evidence:}
    The experiment was designed primarily to observe adversary interaction with the simulated exposure rather than to evaluate the effectiveness of a complete production control stack.
    Control-performance conclusions are therefore limited in this demonstration.

    \item \textbf{Observability evidence:}
    Packet- and application-level telemetry was sufficient to observe request paths, traversal sequences, headers, and interaction with the simulated endpoints.
    The available telemetry was not sufficient to establish actor identity, adversary intent, or deliberate organization-specific targeting.

\end{itemize}

These distinctions prevent the absence or presence of individual observations from being interpreted beyond what the telemetry can support.
In particular, observed exploit-oriented behavior provides evidence of relevance and exposure, but not proof of adversary intent.

\subsection{Updated Judgment and Framework Integration}

The evidence supports the judgment that exploit-oriented probing associated with the modeled FortiWeb management-plane path was relevant to the exposed surface and that the technical preconditions for receiving such traffic existed in the experimental environment.

At the same time, the competing explanation of generic automated Internet scanning cannot be excluded.
The appropriate analytical update is therefore not a claim of confirmed targeted exploitation, but an increase in confidence that the modeled behavior is relevant and locally applicable to publicly exposed FortiWeb-like management surfaces.

Within the FBTM workflow:

\begin{itemize}
    \item A future-oriented safety assumption was made explicit around compromise of an exposed WAF control-plane interface (Step~2);

    \item Attacker-, asset-, and architecture-centric perspectives were used to contextualize adversary relevance, system applicability, required preconditions, and the scenario to be examined (Step~3);

    \item A targeted strategic sensor was deployed and the resulting telemetry was evaluated as adversary, path-feasibility, control-performance, and observability evidence (Step~4);

    \item The resulting judgment can inform governance review concerning exposure of security-control management interfaces, telemetry requirements, and acceptable risk (Step~5); and

    \item Those decisions and any subsequent operational changes can feed revised safety assumptions and future intelligence requirements in later FBTM cycles (Step~6).
\end{itemize}

The case therefore demonstrates the core assumption--hypothesis--evidence sequence of FBTM and illustrates how the resulting evidence can feed governance, operational response, and subsequent learning.
It does not claim to validate the complete framework or demonstrate predictive capability.

\subsection{Case Study Limitations}

Several limitations bound the interpretation of this demonstration.
First, the environment used a simulated FortiWeb management API rather than a production FortiWeb appliance, so the observations do not establish successful exploitation of an authentic vulnerable system.
Second, unsolicited traffic and request patterns cannot by themselves establish actor identity, intent, or deliberate targeting.
Third, the observation primarily evaluates adversary relevance, exposure, and telemetry; it does not constitute a comprehensive evaluation of production control performance.
Finally, a single experimental case cannot establish the comparative effectiveness or empirical validity of FBTM as a whole.

Within these limits, the case illustrates how an explicit Future-Back safety assumption can be translated into a targeted sensor, evaluated through multiple evidence categories, and used to produce a more calibrated security judgment.

\bibliographystyle{unsrtnat}
\bibliography{references}  






\end{document}